\documentclass[preprint,letterpaper]{revtex4}
\usepackage{graphicx}
\def \ta {\theta_a}
\def \tb {\theta_b}

\def \uU {{\underline U}}
\def \ds {\displaystyle}
\def \dsw {D_{{\rm sw}}}
\def \bS {{\bar S}}
\def \vS {{\bf S}}

\def \vQ {{\bf Q}}

\def \vk {{\bf q}}
\def \vq {{\bf q}}
\def \mB {\mu_{\rm B}}

\begin{document}

\title{Spin Dynamics of a Canted Antiferromagnet in a Magnetic Field}
\author{R.S. Fishman}
\affiliation{Condensed Matter Sciences Division, Oak Ridge National Laboratory, Oak Ridge, TN 37831-6032}

\begin{abstract}

The spin dynamics of a canted antiferromagnet with a quadratic spin-wave dispersion near
$\vq =0$ is shown to possess a unique signature.  When the anisotropy gap
is negligible, the spin-wave stiffness $\dsw (\vq ,B) = (\omega_{\vq }-B)/q^2$
depends on whether the limit of zero field or zero wavevector 
is taken first.  Consequently, $\dsw $ is a strong function of magnetic field at a fixed
wavevector.  Even in the presence of a sizeable anisotropy gap, the field dependence of 
the extrapolated $\vq = 0$ gap energy distinguishes a canted antiferromagnet from 
a phase-separated mixture containing both ferromagnetic and antiferromagnetic regions.

\end{abstract}

\maketitle

One of the greatest challenges in magnetism is to identify and characterize a canted 
antiferromagnet (CAF).  Double quantum dots \cite{qd}, cuprates \cite{cup}, 
ruthenates \cite{rut}, RMn$_2$Ge$_2$ compounds \cite{mnge}, Ho and Dy rare-earth 
borocarbides \cite{boro} and intermetallics \cite{intm}, and lightly-doped 
manganites \cite{hir:97,hen:00,geck:01} are all believed to have a CAF phase.  
But in practice, it is extraordinarily difficult to distinguish a CAF from from a 
phase-separated mixture of a ferromagnet (FM) and an antiferromagnet (AF).  This 
Letter demonstrates that a CAF with a quadratic spin-wave (SW) dispersion around 
$\vq =0$ possesses a unique dynamical signature.  In a magnetic field $B$, the SW 
stiffness $\dsw (\vq ,B)=(\omega_{\vq }-\omega_0)/q^2$ of a CAF with negligible 
anisotropy gap approaches different values depending on whether the limit of vanishing 
wavevector or field is taken first.  Consequently, the SW stiffness for a fixed wavevector
changes rapidly in small fields.  Even when the anisotropy gap is sizeable, the field 
dependence of the extrapolated $\vq = 0$ energy gap still distinguishes a CAF from a 
phase-separated mixture containing FM regions.  These results are used to demonstrate 
that the "FM" regions in Pr$_{0.7}$Ca$_{0.3}$MnO$_3$ are actually canted.

The Hamiltonian of a system consisting of spins $\vS_i$ at sites $i$ in a field along 
the $z$ direction can generally be written as $H=H^{(0)}-B\sum_i S_{iz}$ (set $2\mu_B = 1$
until it is needed).  If inversion symmetry is unbroken \cite{dm} and the anisotropy gap is negligible, 
then the small $q$ SW dispersion of a FM or CAF with net magnetization in the $z$ direction can be 
written as 
\begin{equation}
\label{oq}
\omega_{\vq }=\sqrt{\omega_0^2+2\omega_0 E_0 q^2 +D_0^2 q^4},
\end{equation}
where $\omega_0 =B$ is the energy gap and the wavevector $\vq $ lies along one of the crystal axis.  
For simplicity, the lattice constant is set to 1.  

In a FM, the transverse SW frequencies are obtained from the time dependence of 
$S_{i\pm }=S_{ix }\pm iS_{iy}$.  Since $\dot S_{i\pm }=i[H^{(0)},S_{i\pm }]\mp iB S_{i\pm }$, 
the SW frequencies of a FM are simply shifted by $B$.  So for a FM, $D_0=E_0$ and the small $q$ 
dispersion is given by $\omega_{\vq }=B+D_0 q^2$.  For a CAF, the transverse components of the spin 
differ from one site to another and the equilibrium angles depend on field.  Hence,
the above argument fails.  Because a magnetic field does not just shift the SW spectrum, 
it follows quite generally that $D_0\ne E_0$ in a CAF.  This simple conclusion has some 
remarkable consequences.  Notice that $D_0$ and $E_0$ are given by distinct limits of $\dsw (\vq ,B)$:  
$D_0=\lim_{q\rightarrow 0}\lim_{B\rightarrow 0} \dsw (\vq ,B)$ and 
$E_0=\lim_{B\rightarrow 0}\lim_{q\rightarrow 0} \dsw (\vq ,B).$
When the limit of zero wavevector is taken first, $d\omega_{\vq }/dB\rightarrow 1$ but
when the limit of zero field is taken first, $d\omega_{\vq }/dB\rightarrow E_0/D_0 \ne 1$.
At a fixed wavevector $\vq $, Eq.(\ref{oq}) implies that $\dsw (\vq ,B)$ 
is a strong function of field when $B$ is in the neighborhood of $B^{\star } \equiv D_0 q^2$.  For fields 
much less than $B^{\star }$, $\dsw \approx D_0$;  for much larger fields, $\dsw \approx E_0$.

\begin{figure}
\includegraphics *[scale=0.8]{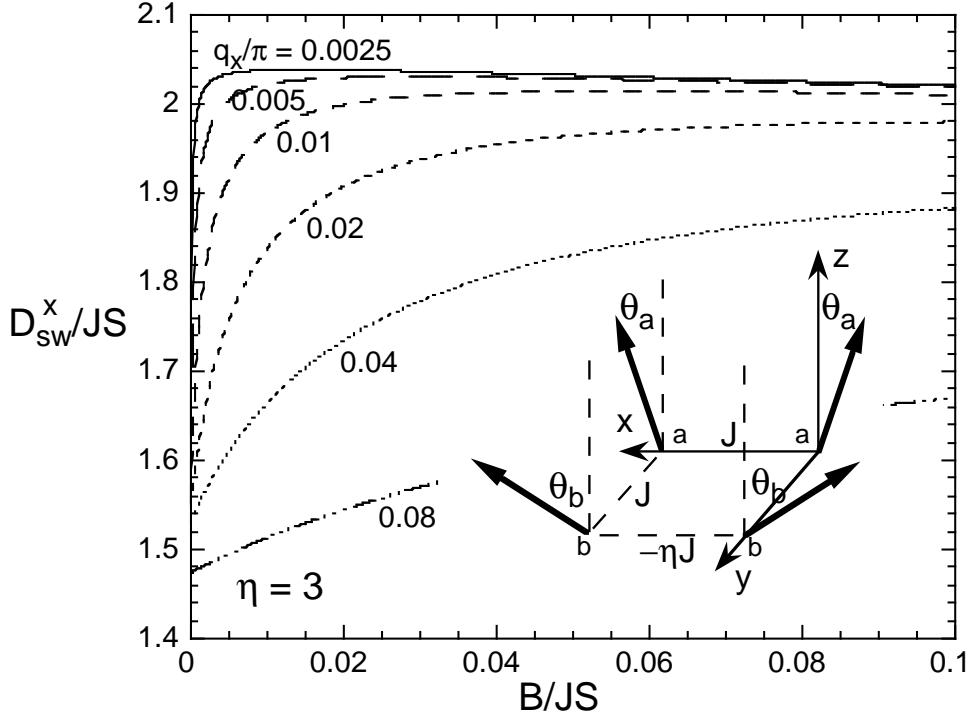}
\caption{
The SW stiffness in the $x$ direction versus field for $q_y=0$ and 
various values of $q_x /\pi $ with $\eta =3$.  Inset is a sketch of the GV model.
}
\end{figure}

To demonstrate these ideas, we consider one of the simplest models for a two-dimensional CAF, which is 
the generalized Villain (GV) model \cite{vil:77,ber:86,gab:89} on a two-dimensional lattice with
three-dimensional spins.  As sketched in the inset to Fig.1, the spins on sublattice $a$ are FM 
coupled to each other and to the spins on sublattice $b$ with exchange constant $J > 0$ while 
the spins on sublattice $b$ are AF coupled to each other with exchange constant $-\eta J$.  
The Hamiltonian of the GV model is $H=-\sum_{\langle i,j\rangle }J_{ij}\vS_i\cdot \vS_j -B\sum_i S_{iz}$, where 
the nearest-neighbor exchange coupling $J_{ij}$ equals either $J$ or $-\eta J$.  The CAF phase is stable when 
$\eta $ exceeds the critical value $\eta_c$, which is $1/3$ in zero field but increases as $B$ increases.  Due to the 
different environments of the $a$ and $b$ sites, the angle $\tb $ at the $b$ sites is always larger than the angle 
$\ta $ at the $a$ sites.

The spin dynamics of the GV model is solved within the rotated reference frame for each spin,  
$\bar \vS_i =\uU_i \vS_i$, where $\uU_i$ is the unitary rotation matrix for site $i$.
A Holstein-Primakoff expansion is performed within each rotated reference frame:
$\bS_{iz} =S-a_i^{\dagger }a_i$, $\bS_{i+}=\sqrt{2S}a_i$, and 
$\bS_{i-} =\sqrt{2S}a_i^{\dagger }$.  
Minimizing the ground-state energy $E=\langle H\rangle $ with respect to $\ta $ and $\tb $ yields 
the relations \cite{gab:89} 
\begin{equation}
\label{v1}
\sin 2\ta +\sin (\ta -\tb )+\frac{B}{2JS}\sin \ta =0, 
\end{equation}
\begin{equation}
\label{v2}
-\eta \sin 2\tb -\sin (\ta -\tb )+\frac{B}{2JS}\sin \tb =0,
\end{equation}
where $BS$ is considered to be of the same order in $1/S$ as $JS^2$.
In zero field, it is easy to show that $\tb =3\ta $ for all $\eta $. 

After expanding $H =E+H_{1}+H_{2}+\ldots $ in powers of $1/\sqrt{S}$, we find that the 
the first-order term $H_{1}$ vanishes provided that the angles $\ta $ and $\tb $ satisfy
Eqs.(\ref{v1}) and (\ref{v2}).  In terms of the Fourier-transformed spin operators
$a_{\vk }^{(r)}$ and $a_{\vk }^{(r) \dagger }$ on the $r=a$ or $b$ sublattice, 
the second-order term can be written as 
\begin{equation}
\label{hv2} 
H_2=JS\sum_{\vk, r,s}\biggl\{ a_{\vk }^{(r)\dagger }a_{\vk }^{(s)}A_{\vk }^{(r,s)}
+\Bigl( a_{-\vk }^{(r)}a_{\vk }^{(s)}+a_{-\vk }^{(r)\dagger }a_{\vk }^{(s)\dagger }\Bigr)
B_{\vk }^{(r,s)} \biggr\},
\end{equation}
with coefficients $A_{\vk }^{(r,s)}$ and $B_{\vk }^{(r,s)}$ given elsewhere \cite{fis:un}.
The Hamiltonian of Eq.(\ref{hv2}) is easily diagonalized \cite{fis:un} using the method originally 
developed by Walker and Walstedt \cite{wal:80} for spin glasses.  The resulting spin-wave 
frequencies in both the CAF and FM phases satisfy the condition $\omega_0=B$;
in the CAF phase, $\omega_{\vQ }=0$ where $\vQ =(\pi ,0)$ is the AF Bragg vector.  
The results of this calculation agree with the SW frequencies numerically evaluated by 
Saslow and Erwin \cite{sas:92}.

In the FM phase with $\eta < 1/3$, the SW stiffnesses are given by the simple expressions 
$D_0^x=E_0^x=(JS/2)(1-\eta )$ and $D_0^y=E_0^y =JS $.  The SW stiffnesses in the CAF phase are derived 
by using Eqs.(\ref{v1}) and (\ref{v2}) to evaluate $d\ta /dB $ and $d\tb /dB$ at zero field and by 
using the SW frequencies \cite{fis:un} to perform a small $q$ expansion of $\omega_{\vq }^2$.  
After a lengthy calculation, we obtain
\begin{equation}
\label{dswx}
D_0^x = \eta D_0^y = JS \sqrt{2}\eta \sqrt{ 1-\ds\sqrt{\frac{\eta }{\eta +1}}},
\end{equation}
\begin{equation}
\label{eswx}
E_0^x = \eta E_0^y = \ds\frac{JS}{2}\ds\frac{3\eta +(1-\eta )\sqrt{\eta /(\eta +1)}}
{\sqrt{2+\sqrt{(\eta +1)/\eta }}}.
\end{equation}
For $\eta > 1/3$, $E_0^{\alpha }> D_0^{\alpha }$ so that the SW stiffnesses are 
enhanced in the limit of small $q$ for fixed field.  The ratio 
$E_0^x/D_0^x=E_0^y/D_0^y$ grows with increasing $\eta $.  In the limit 
$\eta \rightarrow \infty $ as $\ta \rightarrow \pi /6$ and $\tb \rightarrow \pi /2$, 
$E_0^{\alpha }/D_0^{\alpha }\rightarrow \sqrt{\eta /3}$.  Also in the limit of 
large $\eta $, $D_0^x\rightarrow JS\sqrt{\eta }$ diverges but $D_0^y\rightarrow JS/\sqrt{\eta }$ 
tends to zero.

The SW stiffness $\dsw^x(q_x,B)=(\omega_{\vq }-B)/q_x^2$ is plotted versus field in Fig.1 
for $\eta =3$ and for several different values of $q_x/\pi $.  In the limit $q_x\rightarrow 0$ 
for a small but fixed field, $\dsw^x\rightarrow E_0^x\approx 2.05JS$.  But when $B\rightarrow 0$ 
at a small but fixed $q_x$, $\dsw^x \rightarrow D_0^x \approx 1.55JS$.  In practice, 
neutron-scattering measurements in a FM or CAF must avoid the scattering from the lattice Bragg peak 
at $\vq =0$ and the smallest wavevector used to measure the SW frequencies is about 
$0.08\pi $.  For this wavevector, higher-order corrections in $q^2$ contribute to Eq.(\ref{oq}) 
but the SW stiffness in Fig.1 still increases by roughly 15\% as the field increases from 0 to $0.1JS$.  
We emphasize that the dramatic increase in $\dsw^x$ for small fields is {\it not} due to the 
changes in the equilibrium angles $\ta $ and $\tb $, which are minimal, but rather to the general 
inequivalence of $D_0^x$ and $E_0^x$ in a CAF.  However, for very small canting angles 
($\eta $ just above 1/3), $(E_0^{\alpha }-D_0^{\alpha })/D_0^{\alpha }\approx 9\ta^4/8$ so that 
the difference between $D_0^{\alpha }$ and $E_0^{\alpha }$ is proportional to the {\it fourth} 
power of $\ta $ and may not be detectable if the canting angles are too small.

\begin{figure}
\includegraphics *[scale=0.8]{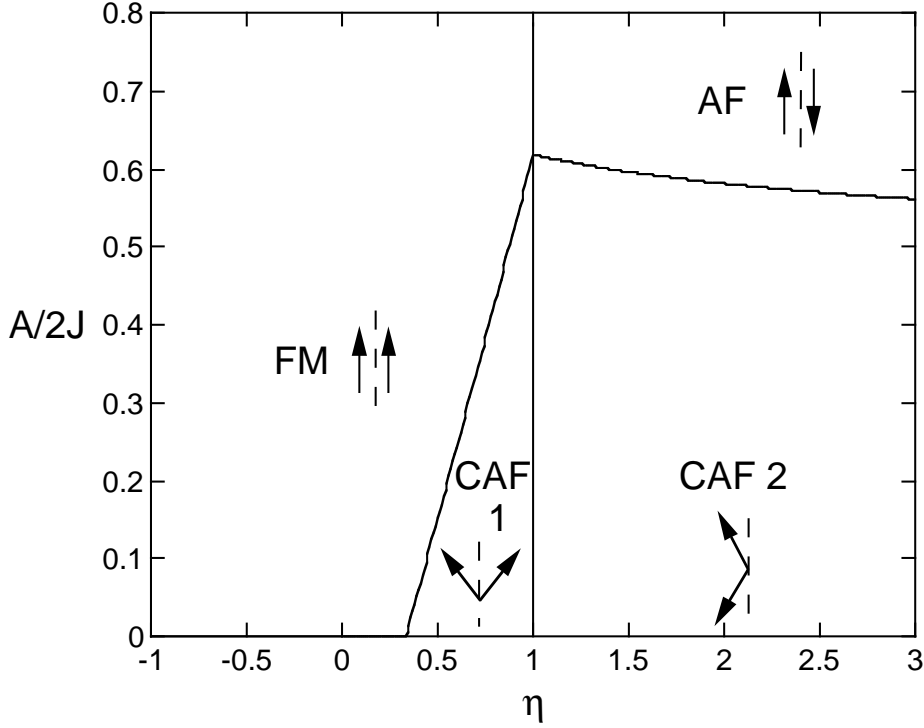}
\caption{
The phase diagram of the GVA model with $B=0$.  Two CAF phases differ in the 
orientation of the spins with respect to the anisotropy axis, which is drawn as the 
dashed vertical line.
}
\end{figure}

Many purported CAF's like the manganites La$_{1-x}$Sr$_x$MnO$_3$ 
and La$_{1-x}$Ca$_x$MnO$_3$ \cite{hen:00} with $0.05 \le x \le 0.125$ have anisotropy 
gaps between 0.2 and 0.5 meV.  To determine the effects of anisotropy on the field dependence 
of the SW stiffness, we add the single-ion anisotropy energy $-A\sum_i S_{iz}^2$ along 
the $z$ axis to the Hamiltonian of the GV model.   Minimizing the energy $E$ of this new 
``GVA'' model in zero field, we obtain the phase diagram in Fig.2.  There are now four 
possible phases: a FM phase for small $\eta $, an AF phase for strong anisotropy and $\eta > 1$,
and two CAF phases.  For $\eta < 1$, the spins in phase CAF 1 are sufficiently aligned 
that the net magnetization points along the anisotropy direction.  For $\eta > 1$, the 
non-colinearity of the spins is large enough that the anisotropy energy is minimized when 
the magnetization lies in the $xy$ plane.  The transition between phases CAF 1 and CAF 2 
is first order with discontinuous changes in $\ta $ and $\tb $.  By contrast, the
transition from CAF 2 to the AF phase is second order, as is the transition from CAF 1 to  
the FM phase.  In a magnetic field applied along the $z$ axis, the spins of the CAF 2 phase will
bend towards the $z$ axis with four inequivalent angles.  While a magnetic field clearly favors 
the CAF 1 phase over the CAF 2 phase, the resulting phase diagram is rather complicated.

An anisotropy gap only appears in the CAF 1 phase.  For the CAF 2 phase, the spins 
and magnetization are still free to rotate about the $z$ axis but rigid rotations about the
direction of the magnetization carry a penalty:  $\omega_0 =0$ but $\omega_{\vQ } > 0$.
This is reversed in the CAF 1 phase:  $\omega_0 > 0$ but $\omega_{\vQ }=0$.  
The harmonic Hamiltonian of the CAF 1 phase maintains the form of Eq.(\ref{hv2}) but 
with revised coefficients.  In the FM phase, the energy gap is given by 
$\omega_0=B+2AS$ and the SW stiffnesses are unchanged.

A difficulty in treating systems with anisotropy is that both the SW stiffness and 
energy gap must be extracted from measurements.  Assuming that two wavevectors $q_1$ 
and $q_2$ are used to fit the form $\omega_{\vq }=\Delta_0 +\dsw q^2$, then the 
extrapolated gap $\Delta_0$ may differ from the true $q\rightarrow 0$ gap $\omega_0$, 
as shown in the inset to Fig.3.  Motivated by measurements on La$_{0.88}$Sr$_{0.12}$MnO$_3$ 
with an anisotropy gap of 0.5 meV and a SW stiffness of 57.5 meV \AA$^2$ \cite{hir:97} 
(the lattice constant is 3.8\AA), we use two sets of parameters to compare the field 
dependence of $\omega_0$ and $\Delta_0$ with the latter averaged over the $x$ and 
$y$ directions.  The wavevectors $q_1=0.1\pi $ and $q_2=0.2\pi $ lie within the range 
of wavevectors used to experimentally extract the energy gap and SW stiffness.  Both sets 
of parameters $\{ A/2J=0.05, \eta =0.61\}$ and $\{ A/2J=0.1, \eta = 0.96\}$ in Fig.3 yield 
the same gap $\Delta_0=0.125 JS$, which gives 0.5 meV for a realistic exchange constant of 
$JS=4$ meV \cite{hir:97}.  For the larger value of $A/2J=0.1$, a higher value of $\eta $ with 
more canted spins ($\ta = 19^o $ and $\tb = 63^o $) is required to produce the same $\Delta_0$ 
as the smaller value of $A/2J =0.05$ ($\ta = 16^o $ and $\tb =50^o $). 

\begin{figure}
\includegraphics *[scale=0.65]{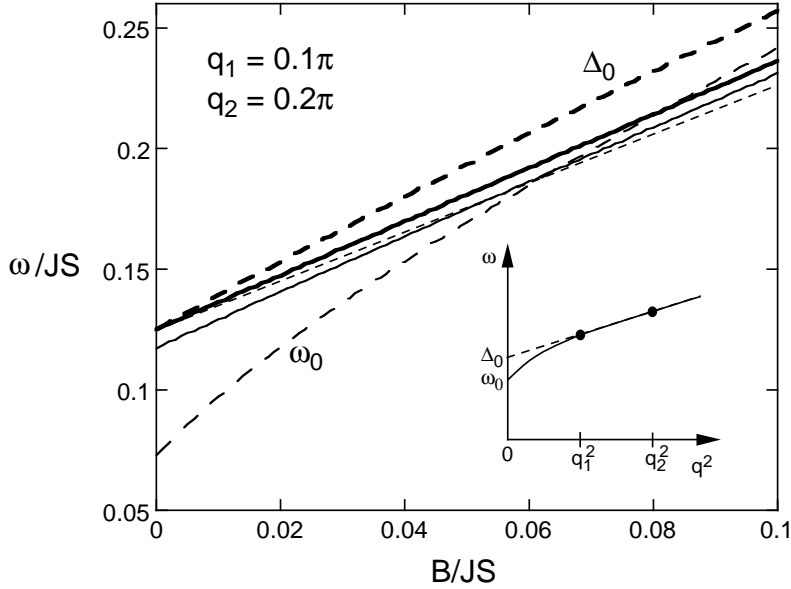}
\caption{
The field dependence of the extrapolated gap $\Delta_0$ (thick curves) and the true 
$\vq =0$ gap $\omega_0$ (light curves), using two values of $q/\pi $ as described 
in the inset and parameters $\{ A/2J=0.05, \eta =0.61\}$ (solid) and 
$\{ A/2J=0.1, \eta = 0.96\}$ (dashed).  The small dash line is the field dependence 
$\Delta_0(B)= \Delta_0(0) +B$ required for a FM. 
}
\end{figure}

Because wavevectors between $q_1$ and $q_2$ fall into the moderate-to-high $q$ limit
with $\dsw q^2 $ comparable to or larger than the energy gap, we may estimate 
$\Delta_0$ by evaluating Eq.(\ref{oq}) in the large $q$ limit:  
$\omega_{\vq }\approx (E_0/D_0) \omega_0 +D_0q^2$, with an extrapolated gap of 
$\Delta_0 \approx (E_0/D_0)\omega_0 $.  For the GVA model parameters in Fig.3, $\Delta_0$ 
overestimates $\omega_0$ by either 6.5 ($A/2J=0.05$) or 71\% ($A/2J=0.1$).  Both $\Delta_0$ 
and $\omega_0$ increase with field as the difference between them diminishes.  But as seen 
in Fig.3, the extrapolated gap $\Delta_0(B)$ increases more rapidly with field than 
it would for a FM.  So if La$_{0.88}$Sr$_{0.12}$MnO$_3$ is really canted, the difference 
(now reinstating $2\mB $) $\Delta_0(B)-2\mB B$ should exhibit significant field dependence.  

\begin{figure}
\includegraphics *[scale=0.65]{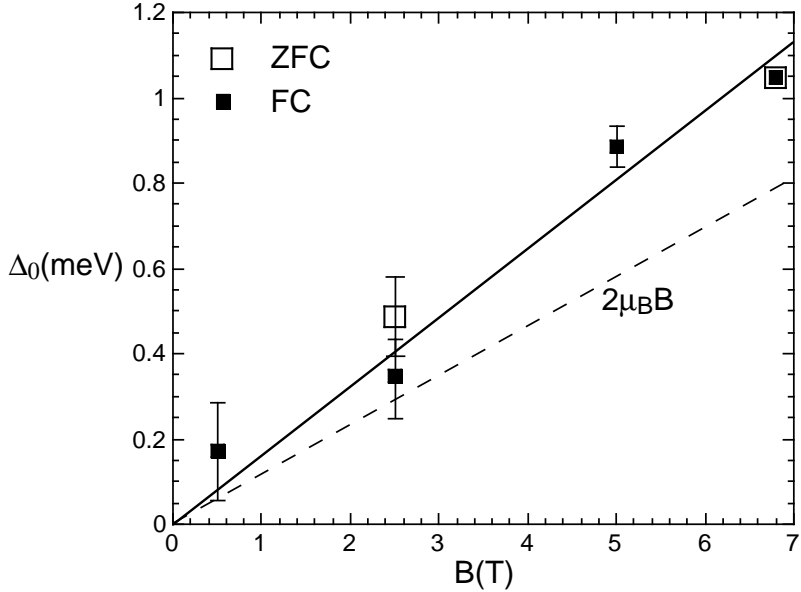}
\caption{
The field dependence of the extrapolated gap for Pr$_{0.7}$Ca$_{0.3}$MnO$_3$ with field 
cooled (FC) and zero-field cooled (ZFC) points shown \cite{fer:02}.  The solid line 
is an interpolation through those points whereas the dashed line is the result $2\mB B$ required 
for a FM.  For 6.8 T, the error bars are smaller than the sizes of the FC and ZFC points.
}
\end{figure}

These results can also be applied to the SW measurements in the low-temperature phase of 
Pr$_{1-x}$Ca$_x$MnO$_3$ with $0.3 \le x \le 0.4$.  Originally believed to be a CAF \cite{jir:85,yos:95}
both above and below the metal-insulator transition at $\sim 3.25$ T, this material is now thought
by some to be phase separated \cite{dea:01,fer:02,sim:02}.  Unlike the manganite discussed above, 
the anisotropy gap is negligible but $\Delta_0(B)$ may still be extrapolated from measurements 
in the large $q$ limit.  For $x=0.3$ \cite{fer:02}, the slope of the extrapolated gap 
$\Delta_0 (B)\approx (E_0/D_0)\omega_0(B)$ plotted versus field in Fig.4 is about 40\% larger than 
the result $\omega_0(B)/B=2\mB $ required for a FM.  While this discrepancy does not gainsay 
the evidence for phase separation in this compound \cite{phs}, we conclude that the "FM" regions in 
Pr$_{0.7}$Ca$_{0.3}$MnO$_3$ must be substantially canted with $E_0/D_0\approx 1.4$.

Other canted systems should be amenable to a similar analysis of the extrapolated energy
gap.  Of particular interest are the CAF phases of the Dy and Ho intermetallics \cite{intm},
which have large moments of over 6 $\mB $ and substantial canting angles.  It would also
be useful to perform this analysis on a wider range of FM materials.  Perhaps because
the result is self-evident, to our knowledge only a single FM material (MnSi above 0.62 T \cite{tar:78}) 
has been studied and shown to obey the required field dependence $\Delta_0(B)=\Delta_0(0)+2\mB B$.

To summarize, we have shown that the field dependence of the SW stiffness and extrapolated energy gap
have unique signatures that distinguish a CAF from a phase-separated mixture containing FM and AF
regions.  Of course, magnetization measurements on single crystals \cite{geck:01}
can also be used to identify CAF's.  But considering the difficulty of those measurements,
the field dependence of the extrapolated energy gap and SW stiffness provide important tools 
to identify and characterize CAF's.  The results of this paper also have important implications 
for comparisons between the predictions of first-principles calculations and experiments, which 
may be describing behavior in different ranges of field and wavevector.
 
It is a pleasure to acknowledge helpful conversations with Drs. W. Saslow, M. Yethiraj, and 
A. Zheludev.  I would especially like to thank J. Fernandez-Baca for sharing his unpublished
data.  This research was sponsored by the U.S. Department of Energy under contract 
DE-AC05-00OR22725 with Oak Ridge National Laboratory, managed by UT-Battelle, LLC.


\begin{references}
\bibitem{phs} The analysis of the extrapolated SW frequencies still holds for a
phase-separated system so long as the smallest $q$ is larger than $2\pi /L$,
where $L$ is the size of the FM or CAF regions.
\end{references}
\end{document}